\documentclass{ifacconf}

\usepackage{graphicx}      
\usepackage{natbib}      
\usepackage{url} 
\usepackage{amsmath,amssymb,amsfonts}
\usepackage{xcolor}
\usepackage{enumitem}
\usepackage{comment}
\makeatletter
\let\old@ssect\@ssect 
\makeatother

\usepackage{natbib}
\usepackage{hyperref}

\makeatletter
\def\@ssect#1#2#3#4#5#6{%
  \NR@gettitle{#6}
  \old@ssect{#1}{#2}{#3}{#4}{#5}{#6}
}
\makeatother
\usepackage{booktabs}
\usepackage{makecell}
\usepackage{multirow}
\usepackage{svg}
\usepackage{tikz}
\usetikzlibrary{automata, positioning, arrows,shapes,positioning,calc}
\usepackage{algorithm}
\usepackage{algpseudocode}

\newtheorem{definition}{Definition}

\newcommand{\mhmargin}[1]{\marginpar{\color{blue}\tiny\tt{MH:} #1}}

\newcommand{\mh}[1]{\textcolor{blue}{#1}}

\newcommand{\tup}[1]{\left( #1 \right)}
\newcommand{\set}[1]{\left\{ #1 \right\}}

\newcommand{\Rn}{\mathbb{R}_{\ge 0}}

\newcommand{\requests}{\mathcal{R}}
\newcommand{\groups}{\mathcal{I}}
\newcommand{\vot}{\beta^\mathrm{vot}}
\newcommand{\nodes}{\mathcal{V}}
\newcommand{\edges}{\mathcal{E}}
\newcommand{\labels}{\mathcal{Z}}
\newcommand{\labelsingle}{z}
\newcommand{\labelsmap}{\mathcal{L}}
\newcommand{\nonnegativenumbers}{\mathbb{R}_{\geq0}}

\begin{document}
\begin{frontmatter}

\title{Invariant Price of Anarchy: a Metric for Welfarist Traffic Control\thanksref{footnoteinfo}} 

\thanks[footnoteinfo]{This work was supported by the ETH Zürich Mobility Initiative (MI-03-22), the ETH Zürich Foundation (2022-HS-213).}
\thanks[footnoteinfo]{This work was supported by the NCCR Automation, a National Centre of Competence in Research, funded by the Swiss National Science Foundation (grant number 51NF40 225155)}

\author[First]{Ilia Shilov\thanksref{equal}}
\author[Second]{Mingjia He\thanksref{equal}}
\author[Third]{Heinrich H. Nax}
\author[Second]{Emilio Frazzoli}
\author[Fourth]{Gioele Zardini}
\author[First]{Saverio Bolognani}

\thanks[equal]{Equal contribution.}

\address[First]{Automatic Control Laboratory, ETH Zurich, 8092 Zurich, Switzerland} 
\address[Second]{Institute for Dynamic Systems and Control, ETH Zurich, 8092 Zurich, Switzerland (e-mail: minghe@ethz.ch)}
\address[Third]{Zurich Center for Market Design \& SUZ, University of Zurich, 8050 Zurich, Switzerland}
\address[Fourth]{Laboratory for Information and Decision Systems, Massachusetts Institute of Technology, Cambridge, MA, USA}

\begin{abstract}  

The Price of Anarchy (PoA) is a standard metric for quantifying inefficiency in socio-technical systems, widely used to guide policies like traffic tolling. Conventional PoA analysis relies on exact numerical costs. However, in many settings, costs represent agents’ preferences and may be defined only up to possibly arbitrary scaling and shifting, representing informational and modeling ambiguities.
We observe that while such transformations preserve equilibrium and optimal outcomes, they change the PoA value. To resolve this issue, we rely on results from Social Choice Theory and define the \emph{Invariant PoA}. By connecting admissible transformations to degrees of \emph{comparability} of agents' costs, we derive the specific social welfare functions which ensure that efficiency evaluations do not depend on arbitrary rescalings or translations of individual costs. Case studies on a toy example and the Zurich network demonstrate that identical tolling strategies can lead to substantially different efficiency estimates depending on the assumed comparability. Our framework thus demonstrates that explicit axiomatic foundations are necessary in order to define efficiency metrics and to appropriately guide policy in large-scale infrastructure design robustly and effectively. 

\end{abstract}


\end{frontmatter}

\section{Introduction}
Socio-technical multi-agent systems, such as transportation networks, energy grids, and multi-robot teams, involve interaction of selfish, heterogeneous agents who jointly determine an equilibrium outcome. 
Because agents optimize individual objectives rather than the system’s optimum, the resulting equilibria are often inefficient. 
The Price of Anarchy (PoA) quantifies this inefficiency as the ratio between the worst-case equilibrium performance and the social optimum~\citep{koutsoupias1999worst, roughgarden2002poa}. 
The PoA has become a key benchmark for designing system coordination policies and utility design~\citep{Paccagnan2022Utility}. 
In transportation, it guides toll design and equilibrium selection to steer networks toward more efficient states~\citep{paccagnan2020utility, chandan2019optimal, chandan2024methodologies, wang2015analysis}.

As PoA measures the ratio between the performance of the resulting equilibrium and that of the social optimum, it inherently depends on the form of the Social Cost Function (SCF) chosen to evaluate this performance. The equilibrium does not change when we apply affine transformations to the agents' costs (such as scaling and shifting, individual or common), as agents' decisions depend on relative differences between states. 
However, the form of the chosen SCF that evaluates the total costs of the system is generally not invariant to such changes. This motivates us to study the \emph{invariance of PoA}, which postulates that \emph{if the equilibrium behavior does not change, then the measure of efficiency of that equilibrium should not change due to informational and modeling ambiguity.} 
To that purpose, we focus on the invariance of the social optimum, using the \emph{welfarist approach}, that allows us to relate the invariance properties of SCF and the chosen affine transformations. 




When individual costs are aggregated through a SCF, usually they are assumed to be measured on the same scale, thus being fully comparable.
However, when agents’ costs can be rescaled or shifted by individual transformations, meaningful aggregation becomes ambiguous.
Social Choice Theory resolves this ambiguity through the \emph{welfarist approach}, which asserts that, under mild assumptions, defining an SCF is equivalent to specifying a degree of \emph{interpersonal comparability} that acts as an informational filter on agents' cost functions~\citep{dAspremontGevers2002}.\footnote{See \cite{shilov2025welfare} for a recent article that formulates the welfaris approach for control problems.} 


The importance of comparability is particularly evident in transportation modeling.
Travelers are heterogeneous: they perceive identical conditions differently.
For instance, the widely used \emph{value of time} converts travel time savings into monetary terms, thereby enabling direct comparison of costs across individuals.
Yet, relying solely on subjective monetary valuation can oversimplify social welfare~\citep{vilain2002differences}.
Moreover, many transportation-related cost components, such as comfort, crowding, perceived safety, and emissions, are difficult to quantify and may be only partially observed or entirely unmeasurable~\citep{goransson2023factors,skoufas2024understanding}.
Even measurable quantities, such as walking or waiting time, often contain biases or measurement errors, introducing arbitrary offsets that distort comparability.

%
%
Finally, the choice of comparability often reflects the planner’s normative stance, given the significant societal implications of transportation systems~\citep{markovich2011social}. 
It encodes what aspects of agents' costs are considered relevant for collective evaluation (e.g., a deliberate modeling decision to exclude external delays occurring outside the road network of interest).

Given these limitations to full cost comparability, it becomes necessary to specify which features of agents' costs should be preserved for welfare evaluation.
Classical results in Social Choice Theory~\citep{sen_interpersonal_1970, roberts_interpersonal_1980, dAspremontGevers2002} formalize this requirement through the concept of  \textit{invariance transformations}.
If the social ranking of outcomes, determined by the SCF, remains unchanged under given class of transformations applied to individual utilities, then that SCF is said to be \textit{admissible} under the corresponding comparability assumption. 
Once the appropriate degree of comparability is specified, the associated efficiency criterion, such as utilitarian, Nash Social Welfare, or max–min, follows naturally.
%


Existing studies on PoA employ various aggregation rules, such as the Nash Social Welfare objective~\citep{branzei2021nash, bilo2022nash} or the Max-Min objective~\citep{koutsoupias1999worst}. 
The utilitarian sum remains the most common formulation in congestion and network routing games~\cite{chandan2019optimal}, yet it implicitly assumes exact cardinal comparability of agents' costs and is therefore not invariant to individual affine transformations. 
The generalized PoA proposed in~\citep{chandan2024methodologies} achieves invariance under \emph{common} affine transformations but not under \emph{individual} ones.
In contrast, our framework provides a unified treatment of PoA invariance by aligning the choice of the SCF with the invariance properties of the game's solution concept, thereby ensuring that the resulting efficiency metric remains robust to the numerical representation of agents' preferences.


The main contribution of this paper is conceptual. 
We develop a rigorous axiomatic framework for evaluating system efficiency under strategic interaction, formally defining the \emph{Invariant Price of Anarchy} by linking the game-theoretic solution concepts with the possibility results of Social Choice Theory. The framework is demonstrated on both a toy network example and a case study of the Zurich (Switzerland) transportation system. Our results show that depending on the assumed level of interpersonal comparability, identical tolling strategies can lead to substantially different efficiency assessments. This shows that explicit axiomatic foundations are essential for defining robust efficiency metrics and for avoiding misguided policy conclusions in large-scale infrastructure design.

\section{Motivation}\label{sec:motivation}

Invariance to admissible transformations can come from two sources: \emph{informational invariance} and \emph{modeling invariance}.
Informational invariance concerns what is \emph{known} about the representation of players' preferences. In addition, the modeler makes deliberate choices about \emph{what to include} in the model, for example by accounting for fixed base delays or by rescaling costs into normalized units. Such choices leave the feasible outcomes and the incentives that determine equilibrium unchanged, yet the standard PoA can be sensitive to them. We illustrate these properties on the simple examples from transportation modeling.


\emph{Offset invariance.}
\begin{figure}[t]
    \centering
    \includegraphics[width=0.7\linewidth]{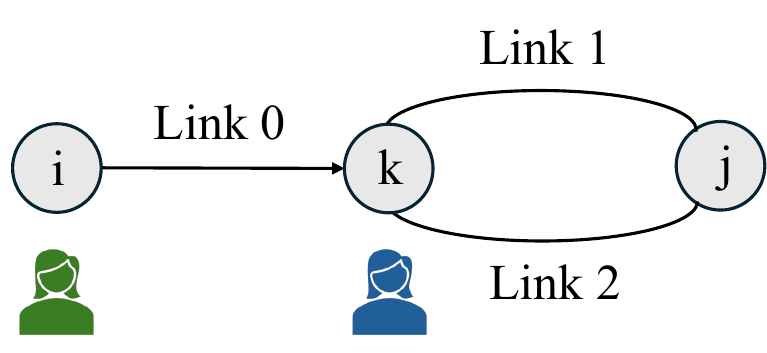}
    \caption{Parallel network in which link \(0\) represents an exogenous access segment.}
    \label{fig: small_network}
\end{figure}
Consider the network in Figure~\ref{fig: small_network}. Agents starting at node \(i\) must traverse link \(0\) (cost \(b_0\)) before choosing between links \(1\) and \(2\).
The cost on route \(k\) is \(c_{k}(x)=b_0+\ell_k(x_k)\). 
Because Wardrop equilibria depend only on differences in route costs, adding the constant \(b_0\) does not affect equilibrium flows. 
In contrast, a utilitarian social cost aggregates all individual costs and therefore shifts by exactly \(b_0\). By increasing \(b_0\) one can move equilibrium and optimal welfare levels arbitrarily closer without altering behavior on links \(1\) and \(2\). 
Hence, a PoA defined as a ratio of cost levels becomes sensitive to arbitrary modeling choices about access segments that policy cannot influence.
This reasoning extends to agent-specific offsets \(b_i\), reflecting, for instance, delays accumulated outside the modeled network. An efficiency measure should therefore ignore modeling components that are constant across all feasible outcomes, whether the constant is common to all agents or origin specific, as in the example above.
In particular, the PoA that captures the inefficiency of the parallel network between \(k\) and \(j\) should not depend on whether a traveler starts at node \(i\) or at node \(k\).

\emph{Scaling invariance.}
In many systems the choice of cost scale is part of the model specification. 
Travel time may be expressed in different units, and in load balancing or scheduling it is common to normalize costs by a maximal load~\citep{bilo2022nash}. 
In such cases, PoA should be invariant to multiplicative rescalings of costs.
Otherwise, the efficiency assessment can be distorted by an arbitrary unit choice or by a single very large normalized entry.
From this perspective, the Nash Social Welfare (NSW) is a natural aggregation rule because its geometric-mean structure is scale-neutral for positive rescalings~\citep{FlemingWallace1986HowNotToLie}.


In its standard form, NSW is defined relative to a disagreement outcome \(r\) and maximizes the product of utility gains. 
In a cost-minimization setting, this translates into maximizing the product of cost savings relative to a reference cost \(c^{max}\): \(\prod_{i}(c^{max}_i - c_i(x))\). 
The choice of reference is application-dependent: it can represent a status quo~\citep{BinmoreRubinsteinWolinsky1986NBS}, an outside option, a minimum service level, or a baseline allocation~\citep{kaneko1979nash}. 
Combined with the requirement of \emph{offset invariance}, this perspective suggests expressing PoA comparisons in terms of \emph{cost differences} relative to an appropriate reference, a principle that we formalize in the next section.

\section{Welfarist Framework for Price of Anarchy}\label{sec: preliminaries}

Under the welfarist framework, \emph{comparability assumption} allows us to explicitly specify which features of costs (levels, scales, or only order) are meaningful across agents, and therefore which transformations leave welfare evaluation invariant.
We use this perspective to classify games by their underlying comparability and to derive the associated admissible aggregations (e.g., utilitarian, NSW, max-min).
In doing so, PoA becomes invariant to representation choices consistent with the declared comparability, ensuring a stable and policy-relevant efficiency metric.

\subsection{Preliminaries}
We consider a system with heterogeneous agents, modeled as a strategic game with players~$\mathcal{N} = \{1,\ldots,n\}$. 
Each player \(i \in \mathcal{N}\) is equipped with a set of strategies \(S_i\), and \(c_i : \mathcal{S} \to \mathbb{R}\) denotes the cost function for player \(i\). The joint strategy profile is denoted by $s=(s_1, s_2, \ldots, s_n)$, where $\mathcal{S} = \prod S_i$. 
Each player's cost function $c_i: \mathcal{S} \to\mathbb{R}$ maps from a strategy profile to the cost borne by player $i$. 
A common (often implicit) assumption in the literature is that the family~$\{c_i\}_{i\in\mathcal{N}}$ is \emph{fully comparable} across agents, i.e., both \emph{levels} and \emph{increments} of cost admit meaningful interpersonal comparison.
We relax this assumption in the following subsections.

System performance is measured by a SCF~$C_{\text{sys}} : \mathcal{S} \to \mathbb{R} $, which assigns a total cost to each~$s \in \mathcal{S}$. 
A system-optimal strategy profile satisfies $s_{\mathrm{opt}} \in \arg\min_{s \in \mathcal{S}} C_{\text{sys}}(s)$. 
Thus, a cost-minimization game describing the system is represented by the tuple
\begin{equation}\label{eq: game_original}
    G:= \{\mathcal{N}, \mathcal{S}, \{c_i\}_{i \in \mathcal{N}}, C_{\text{sys}}\}
\end{equation}
Throughout, we adopt the \emph{welfarist approach}~\citep{roberts_interpersonal_1980}, as adapted to control applications in \cite{shilov2025welfare}: under mild conditions, the social objective is a function of the individual costs. 
This approach is widely employed in the literature, most commonly through the \emph{utilitarian} specification, i.e. the social cost is defined by summing individual costs $C_{\text{sys}}(s) = \sum_{i=1}^n c_i(s)$~\citep{roughgarden2002poa, roughgarden2004bounding, chandan2024methodologies}.


\subsection{Price of Anarchy}

While the optimal allocation minimizes the SCF, in a game theoretic framework the system performance might not coincide with this optimum due to the selfish behavior of the players. 
To quantify the inefficiency induced by decentralized, self-interested decisions, we focus on pure Nash equilibria (PNE). 
A strategy profile $s^*$ is a PNE if no player can unilaterally deviate to reduce their cost, that is, for every $i\in \mathcal{N}$, $c_i(s_i^*,s_{-i}^*) \le c_i(s_i,s_{-i}^*)$ for all $s_i\in S_i$. 
Let $\mathrm{NE}(G)$ denote the set of all PNE of game $G$. 

For a given SCF, the optimal social cost is $C^\star=\min_{s\in \mathcal{S}} C_{\text{sys}}(s)$, while the worst case equilibrium cost is $C_{NE}=\max_{s\in \mathrm{NE}(G)} C_{\text{sys}}(s)$. The standard \emph{Price of Anarchy} is then $\mathrm{PoA}(G)=C_{NE} / C^\star \ge 1$.

\emph{Invariant PoA}
As argued in Section~\ref{sec:motivation}, modeling choices such as a common offset in all costs can arbitrarily distort standard efficiency ratios. To ensure the metric depends only on the game's structure (informational invariance) and not on arbitrary base costs, we must measure efficiency relative to a meaningful boundary condition. 

Let $c^{\max} = (c^{\max}_1, \dots, c^{\max}_n)$ be a vector of \emph{reservation costs}, where $c^{\max}_i$ represents the threshold cost at which agent $i$ withdraws from the system (the participation constraint). It represents the ``breakdown state'' of the system or the value of the outside option (e.g., not traveling). By individual rationality, agents only participate if their cost is strictly below $c^{max}_i$. Thus, for any feasible profile $s \in \mathcal{S}$ with active agents, the condition $c^{\max}_i > c_i(s)$ is naturally satisfied.
The \emph{cost saving} (or surplus) for agent $i$ is defined as:
\begin{equation}
    \Delta c_i(s) = c^{\max}_i - c_i(s).
\end{equation}

Since the cost saving $\Delta c_i(s)$ is a strictly decreasing affine transformation of the cost $c_i(s)$, maximizing individual surplus is equivalent to minimizing individual cost. Consequently, the set of Nash Equilibria and the system optimum remain invariant whether agents optimize $c_i$ or $\Delta c_i$.

Accordingly, we define the system welfare $W_{\Delta}(s)$ as an aggregation of the cost savings vector $\Delta c(s) = (\Delta c_1(s), \dots, \Delta c_n(s))$. The discussion on the specific form for $W_{\Delta}(s)$ is laid out in Section \ref{sec:welfarism}.
The \emph{Invariant PoA} is defined as the ratio of the optimal potential welfare (surplus)  to the worst-case equilibrium welfare:
\begin{equation}\label{eq:poa_reference}
    \mathrm{PoA}_\Delta(G, c^{\max})=\frac{\max_{s\in \mathcal{S}}\,W_{\Delta}(s)}{\min_{s\in \mathrm{NE}(G)}\,W_{\Delta}(s)}.
\end{equation}
This formulation, coupled with comparability assumptions introduced in the next subsection, ensures that the efficiency metric remains stable under affine transformations of the underlying costs, satisfying the invariance principles established in Section~\ref{sec:motivation}.


\subsection{Welfarism}\label{sec:welfarism}
The welfarist approach guides the choice of an aggregation function $W_{\Delta}(s)$ consistent with the invariance properties. 
Once the individual costs and the \emph{reservation profile} are fixed, all information relevant for a social decision is contained in the vector of surpluses at each outcome. 

Given a surplus profile~$\Delta c(s)$, a social preference~$\succeq$ ranks strategy profiles (outcomes)~$x, y \in \mathcal{S}$. 
Classic results show that, under mild conditions, such a ranking is represented by a continuous social welfare function \(W:\mathbb R^n\to\mathbb R\), and that the admissible form of \(W\) is determined by the degree of interpersonal comparability \citep{roberts_interpersonal_1980, dAspremontGevers2002, Sen1979_welfarism, shilov2025welfare}. 
For the social preference to be consistent, we first specify which \emph{fundamental properties (or axioms)} it needs to satisfy: Weak Pareto (WP), Partial Independence of Irrelevant Alternatives (PI), and Continuity (C), stated here informally. 
We refer to~\cite{shilov2025welfare} for a formal treatment and discussion.

\begin{description}
    \item[WP] For any two profiles $x, y \in \mathcal{S}$, if \(\Delta c_i(x) > \Delta c_i(y)\) for all \(i\in \mathcal{N}\), then \(x\succ y\).
    \item[PI] \cite[Sec. 5]{roberts_interpersonal_1980}. Once the reservation costs and the surpluses on the compared set are fixed, the ranking on that set does not depend on costs at other points.
    \item[C] Small changes in \(\Delta c(x)\) and \(\Delta c(y)\) do not change a strict ranking.
\end{description}
An optional property that one might want to impose is Anonymity (A), which requires equal treatment of agents who differ only by labels.
It can be seen as a desirable fairness property, but if predetermined priorities are to be incorporated, anonymity may not be appropriate.
\begin{description}
    \item[A] For any permutation \(\pi: \mathbb{N} \to \mathbb{N}\) of agents, the ranking given by $W$ remains the same.
\end{description}

Under WP, PI, and C there exists a continuous \(W:\mathbb R^n\to\mathbb R\) such that
\[
x\succeq y\ \Longleftrightarrow\ W_{\Delta}(x)\,\ge\,W_{\Delta}(y),
\]
i.e., the social ranking depends only on cost savings relative to the reservation profile. 
The exact form of \(W\) is determined by the chosen level of comparability. In what follows we focus on the cardinal classes that matter for numerical efficiency evaluation, following \cite{roberts_interpersonal_1980}.


\subsection{Comparability classes by invariance} \label{section: Comparability classes by invariance}
A social welfare function~$W_{\Delta}(s):\mathbb R^n\to\mathbb R$ ranks outcomes using the reference-adjusted quantity. 
The exact shape of $W_{\Delta}$ is constrained by the information the designer accepts about interpersonal comparison.
We encode that information through an invariance requirement.

\begin{figure}[tb]
    \centering
    \includegraphics[width=80mm]{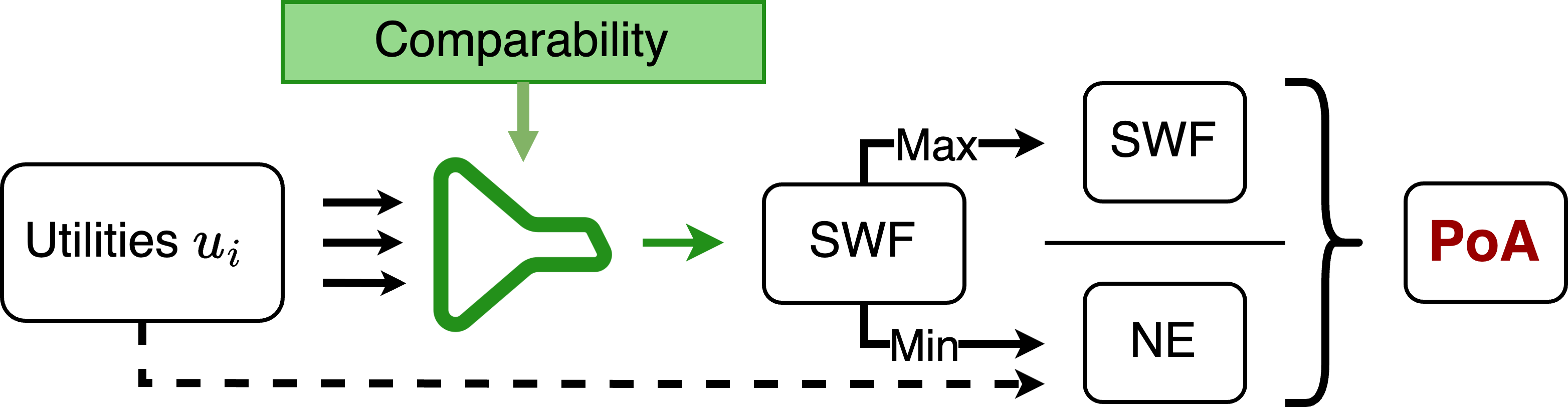}
    \caption{Inclusion of Comparability for Surplus Maximization Games}
    \label{fig: scheme_introduction}
\end{figure}

\begin{definition}\label{def:invariance}
Let $\Phi$ be a family of strictly increasing real maps and fix a reservation profile $c^{\max}$. 
A list $\varphi=(\varphi_1,\dots,\varphi_n)$ with $\varphi_i\in\Phi$ is an \emph{invariance transformation} if for all $x,y\in \mathcal{S}$,
    \begin{equation*}
        W_{\Delta}(x)\ \ge\ W_{\Delta}(y)
         \Leftrightarrow
        W_{\Delta}^{\varphi}(x)\ \ge\ W_{\Delta}^{\varphi}(y),
    \end{equation*}
    where transformed surplus is $\Delta c_i^{\varphi}(s)=\varphi_i(c^{max}_i)-\varphi_i(c_i(s))$ and $W_{\Delta}^{\varphi}(s)=W\!\left(\Delta c^{\varphi}(s)\right)$.
\end{definition}

The choice of $\Phi$ specifies what is admissible for interpersonal comparison and leads to different canonical forms for $W_{\Delta}$ \citep{shilov2025welfare}. 
We focus on two main cases.

\subsubsection{Cardinal Non Comparability (CNC)}
Here $\Phi_{CNC}$ contains all agent-specific positive affine maps $\varphi_i(c_i)=a_i c_i + b_i,\quad a_i>0.$ Units and zeros may differ across agents, corresponding to the most general setting. 
The only meaningful statements compare ratios of each agent's savings relative to the reservation baseline. 
Given the reservation profile $c^{\max}$ (s.t. $c_i(s) < c^{\max}_i\ \forall i \in \mathcal{N}$, \cite{shilov2025welfare}) and exclusion of indifferent agents ($c_i(s^{\ast}) = c^{\max}_i$) at that point, the ranking is given by the Nash SWF:
\begin{equation*}
    W_{\Delta}(s)\;=\;\prod_{i \in \mathcal{N}} \Delta c_i(s)^{w_i}
\end{equation*}
with $w_i>0$ and $w_i = w_j$ if \textbf{A} is imposed. This form maximizes the geometric mean of savings.

\subsubsection{CUC: Cardinal Unit Comparability}
Here $\Phi_{\mathrm{CUC}} = \big\{\,\varphi_i(t)=a\,t+b_i\ \text{with}\ a>0,\ b_i\in\mathbb R\,\big\}.$ Then $\Delta c^{\varphi}_i(s)=a\,\Delta c_i(s)$ for every $i$, meaning that while base costs differ, the \emph{scale} of cost units is common across agents (e.g., currency). Under WP, PI, and C with the reservation profile $c^{\max}$, the admissible reference adjusted welfare has the isoelastic (Atkinson) form:
\[
W_{\Delta}(s)
=\begin{cases}
\displaystyle \sum_{i\in \mathcal{N}} w_i\,\frac{\big(\Delta c_i(s)\big)^{\,1-\rho}}{1-\rho}, & \rho\neq 1,\\[1.2ex]
\displaystyle \sum_{i\in \mathcal{N}} w_i\,\log\big(\Delta c_i(s)\big), & \rho=1,
\end{cases}
\qquad w_i>0.
\]
The parameter $\rho \ge 0$ encodes the planner's \emph{inequality aversion} regarding the distribution of savings:
\begin{itemize}
    \item \textbf{Utilitarian ($\rho=0$):} Maximizes the weighted sum of savings $\sum w_i \Delta c_i(s)$. This is equivalent to minimizing the weighted sum of costs (standard efficiency).
    \item \textbf{Nash ($\rho \to 1$):} Maximizes the sum of logarithmic savings. Note that this is order-equivalent to the Nash Product, recovering the CNC ranking.
    \item \textbf{Max-Min ($\rho \to \infty$):} Maximizes the minimum saving $\min_i (\Delta c_i(s))$. This prioritizes the agent closest to reservation cost, regardless of total system efficiency.
\end{itemize}

In what follows, we focus on these three aggregators as representative points along the CUC spectrum. 
Each arises either from a particular stance on $\rho$ (see, e.g., \citep{lan_fairness}) or from simple additional axioms (see \cite{BossertKamaga2020MixedUtilitarianMaximin}). 
A full axiomatization and a comparison across intermediate $\rho$ values are beyond the scope of this work.
\section{Welfarist Traffic Control}\label{sec: traffic model}
Building on the invariance-based welfarist framework, we now instantiate the approach in a traffic-routing context. 
The goal is to (i) model traveler behavior and a planner’s interventions on a transportation network, (ii) derive system-optimal assignments under a chosen comparability class, and (iii) evaluate efficiency via the reference-adjusted PoA.

\subsection{Problem setting}
\subsubsection{Transportation network}
Consider a transportation network represented by a directed graph $G = (\nodes, \edges, \labelsmap)$,
where $\nodes$ is the set of vertices, $\edges \subseteq \nodes \times \nodes$ is the set of directed edges, and $\labelsmap: \edges \to  \labels $ is a mapping from the set of edges $\edges$ to the set of edge labels $\labels$.
Each edge $e \in \edges$ carries a label $\labelsingle_e = (x_e, t_e, \pi_e) \in \labels = \nonnegativenumbers^3$,  where 
$x_e$ is the link flow, $t_e$ is the edge travel time and $\pi_e$ is the monetary cost associated with edge $e$.
The travel time function $t_e:\nonnegativenumbers \rightarrow \nonnegativenumbers$ is a mapping from the edge flow $x_e$ to the corresponding edge travel time $t_e(x_e)$. 
A widely used specification is the Bureau of Public Roads (BPR) function \citep{BPR}, which is strictly monotone and continuous in the link flow, and is given by
\begin{align}
t_e(x_e) = t_e^0 \tup{1 + a \tup{\frac{x_e}{C_e} }^{b}}, \nonumber
\end{align}
where $t_e^0$ denotes the free-flow travel time on edge $e$, 
$C_e$ is the capacity of the edge, 
and $a=0.15, b =4$ are empirical parameters.

\subsubsection{Heterogeneous travelers}
We define a heterogeneous set of traveler groups, indexed as $i \in \groups = \set{1,2,3}$, 
where $i=1$ corresponds to business travelers, $i=2$ to commuting travelers, and $i=3$ to leisure travelers. 
Different groups perceive traffic-related factors differently: business travelers prioritize time and are relatively
insensitive to monetary cost; commuters exhibit intermediate sensitivities; leisure travelers are more flexible in time but highly cost-sensitive. 
Each request is represented by a vector $r_i \in \requests_i= \tup{o, d, n^{od}_i},$ 
where $(o,d) \in \nodes \times \nodes$ is an OD pair, $i \in \groups$ is the traveler group, 
and $n^{od}_i \in \Rn$ is the number of requests from group $i$ traveling from $o$ to $d$.  
The total number of requests from traveler group $i$ is then given by $n_i=\sum_{r_i \in \requests_i}n^{od}_i$.
The travel demand between origin $o$ and destination $d$ will be extracted from data and is fixed.
For each OD pair $(o,d)$, let $P_{od} \subseteq \edges$ denote the set of all feasible paths connecting $o$ to $d$.  



\subsubsection{Utility representation}
The generalized cost (disutility) perceived by a traveler in group~$i$ when 
selecting a path $p \subseteq \mathcal{E}$ is defined as
\begin{align}
    c_i(p)
    = \sum_{e \in \mathcal{E}} 
    \delta_{e}^{p}\,\bigl( \vot_i\, t_e(x_e) + \pi_e \bigr),
    \label{user_disutility}
\end{align}
where $\vot_i$ denotes the value of time for group~$i$, 
and the indicator $\delta_{e}^{p}$ equals $1$ if edge~$e$ lies on path~$p$ and 
$0$ otherwise.
We use $U_i$ to denote the average travel utility of traveler group $i$, defined as
\begin{align}
    U_i= c_i^{\max}- \frac{1}{n_i} \sum_{e \in \edges}  x_e^i \tup{\vot_i \, t_e(x_e) + \pi_e }
\end{align}
where $c_i^{\max}$ represents the reservation cost, i.e., the maximum cost a traveler is willing to incur; travelers opt out of the routing game if their cost exceeds this value. Similar participation considerations have been incorporated in prior travel-behavior studies \citep{golob1981utility, bowman2001activity}. In this work, we focus on the case in which travelers do not have convenient alternatives to traveling, so $c^{max}_i$ is large and $U_i$ is ensured to be positive. Future research will focus on multi-modal mobility systems where travelers have a lower reservation cost because they can access other alternatives.

\subsection{Equilibrium flow with self-interested travelers}
If each traveler is assumed to freely select a path that minimizes their perceived generalized cost, 
according to Wardrop’s First Principle \citep{wardrop1952}, no traveler can reduce their travel cost by unilaterally changing routes at a Nash equilibrium.
Based on this principle, the equilibrium flow can be characterized as the solution to the following optimization problem:
\begin{subequations}
\begin{align}
\max_{x} \quad & -\sum_{e \in \edges} \int_{0}^{x_e} t_e(w) \, dw 
- \sum_{e \in \edges} \sum_{i \in \groups} \frac{x^i_e}{\vot_i} \pi_e \\
\text{s.t.} \quad & \sum_{p \in p_{od}} f_p^{i,od} = n_i^{od}, \quad \forall (o,d) \in (\nodes,\nodes), i \in \groups, \label{eq:f_a}\\
& f_p^{i,od} \geq 0, \quad \forall p \in p_{od}, (o,d) \in (\nodes,\nodes), i \in \groups, \label{eq:f_0}\\
&x^i_e = \sum_{o \in \nodes} \sum_{d \in \nodes}  \sum_{p \in p_{od}}  \sum_{i \in \groups} \delta_{e}^p f_p^{i,od}, \quad \forall e \in \edges, \label{eq:x_a}\\
&x_e =  \sum_{i \in \groups} x^i_e, \quad \forall e \in \edges, \label{eq:x_i}
\end{align}
\end{subequations}
where 
$f_p^{i,od}$ represent the flow of traveler group $i$ on path  $p \in p_{od}$  for OD pair $(o,d)$. 
Constraints \eqref{eq:f_a} ensure that all requests are satisfied.
Constraints \eqref{eq:x_a} and \eqref{eq:x_i} ensure the flow conservation with heterogeneous traveler groups.

\subsection{Welfarist traffic control}
\subsubsection{Invariant price of anarchy}

As discussed in Section~\ref{sec:welfarism}, different comparability assumptions lead to different admissible formulations of PoA, given as follows:
\begin{align}
    \mathrm{PoA}^\mathrm{CUC_0} & = \frac{\sum_{i \in \mathcal{I}} n_i U^*_i}{\sum_{i \in \mathcal{I}} n_i U^{NE}_i}, \label{eq:poa_cfc}\\
    \mathrm{PoA}^\mathrm{CNC} & = \frac{\prod_{i \in \mathcal{I}} n_i U^*_i}{\prod_{i \in \mathcal{I}} n_i U^{NE}_i}, \label{eq:poa_cnc}\\
    \mathrm{PoA}^\mathrm{CUC_{\infty}} & = \frac{\min_{i \in \mathcal{I}} U^*_i}{\min_{i \in \mathcal{I}} U^{NE}_i},\label{eq:poa_olc}
\end{align}
where $U^*_i$ and $U^{UE}_i$ denote the average utility of traveler group $i$ under the system-optimal and traveler equilibrium assignments, respectively, and $n_i$ is the total demand of group $i$.  
Under the $CUC_0$ assumption, \eqref{eq:poa_cfc} measures PoA when travelers’ utilities $U_i$ are assumed to be fully comparable in both level and scale, with parameter $\rho :=0$ 
\eqref{eq:poa_cnc} is based on assuming that only relative differences in travelers’ utilities are comparable, enabling aggregation through a demand-weighted geometric mean.
In contrast, \eqref{eq:poa_olc} evaluates system efficiency by focusing on the most disadvantaged traveler group, i.e., the group with the lowest travel utilities, obtained by taking $\rho \rightarrow \infty$, as in Section \ref{sec:welfarism}. For a more detailed discussion on the choice of comparability assumption, see \citep{shilov2025welfare}.

\subsubsection{Utility aggregation and optimal traffic assignment}
The system-optimal traffic assignment framework assumes there is a central planner allocating routes to all travelers with the objective of maximizing overall system welfare.
The optimal assignment produces the set of optimal utilities~$(U^*_i)_{i \in \mathcal{I}}$.
The general form of the system-optimal design problem is given by:
\begin{align}
\max_{x} \quad & W(x) \\
\text{s.t.} \quad & \eqref{eq:f_a}-\eqref{eq:x_i}, 
\end{align}
where the feasible region is defined by demand conservation and nonnegativity constraints.

The formulation of a socially optimal traffic assignment critically depends on how travelers’ preferences are assumed to be comparable across traveler groups, which will lead to different specifications of the social welfare function $W(x)$, and consequently to different optimization problems. 
With the CUC$_0$ assumption, utilities are assumed to be comparable both in scale and in level, which justifies a utilitarian aggregation across groups. 
This corresponds to maximizing
\begin{align}
    W^\mathrm{CUC_0}(x^*) = \sum_{i \in \mathcal{I}}\sum_{e \in \mathcal{E}} c_i^{\max}- x_e^i  \tup{\vot_i \, t_e(x_e) + \pi_e }
\end{align} 
If only relative differences in utilities are regarded as meaningful, then the CNC assumption applies. In this case, social welfare is represented by a demand-weighted geometric mean,
\begin{align}
    W^\mathrm{CNC}(x^*) = \prod_{i \in \mathcal{I}} \tup{c_i^{\max}- \sum_{e \in \mathcal{E}} x_e^i  \tup{\vot_i \, t_e(x_e) + \pi_e }},
\end{align}

Finally, under the CUC$_{\infty}$ assumption, utilities cannot be aggregated but can be ordered. The social planner will maximize the utility of the most disadvantaged group:
\begin{align}
    W^\mathrm{CUC_{\infty}}(x^*) = \min_{i \in \mathcal{I}} \sum_{e \in \mathcal{E}} \tup{c_i^{\max}- x_e^i  \tup{\vot_i \, t_e(x_e) + \pi_e }}.
\end{align}

\subsection{Algorithmic solution}
When adopting selected comparability assumptions $\Phi$, we obtain the corresponding
social welfare function $W^{\Phi}$ to be maximized. 
To search for optimal solutions for the traffic assignment problem, we adopt the Frank–Wolfe algorithm. 
Algorithm~\ref{alg:fw} summarizes the iterative procedure used to compute the system-optimal 
traffic assignment under a given comparability structure~$\Phi$.  
Let $M^\Phi: \nonnegativenumbers^{|\edges|} \rightarrow  \mathbb{R}^{|\edges| \times |\groups|}$ denote the function mapping from the current flow into marginal edge welfares for traveler groups.
At each iteration $t$, the algorithm evaluates the welfare margins 
$w' = M^{\Phi}(x^{t})$ given the current flow $x^{t}$. 
These margins are then used to construct an auxiliary flow $y^{t}$. 
A stepsize $\gamma$ is then calculated to update the flow based on the current and auxiliary 
flows so as to maximize the welfare along the search direction.
The final iterate $x^{*}$ returned by the algorithm represents the system-optimal solution consistent with the chosen comparability assumption~$\Phi$.

%
The specific form of the welfare objective $W^{\Phi}$ determines how utilities of
traveler groups are aggregated; consequently, the marginal welfare functions
$M^{\Phi}_{i}$ differ across the chosen comparability assumtion $\Phi$. 
Under $\text{CUC}_0$, when the goal is to maximize the sum of travel utilities, the objective function is differentiable.
\begin{align}
    M^{\mathrm{CUC_0}}_{i}(x_e)
= -\tup{\vot_i t_e(x_e)+\pi_e + t'_e(x_e)\,\sum_{h\in\mathcal{I}} x_e^h \vot_h},
\nonumber
\end{align}
where the first term represents the direct marginal impact on group $i$ and the second term represents the impact on all the traveler groups. 
$t'_e(x_e)$ denotes the derivative of the travel time function $t_e$ with respect to the edge flow $x_e$.

When the aim is to maximize the product of groups’ utilities, a scaled subgradient is given by:
\begin{align}
    M^{\mathrm{CNC}}_{i}(x_e)
    \;=-\tup{\; \frac{\vot_i\,t_e(x_e)+\pi_e}{C_i}
    \;+\; t'_e(x_e)\,\sum_{h\in\mathcal{I}} \frac{x_e^h\,\vot_h}{C_h} \,}, \nonumber
\end{align}
where the first term is the direct effect for group $i$ normalized by the group total cost $C_i$; the second term is the cross-group coupling weighted by $1/C_h$.

When maximizing the utility of the least-advantageous group, the subgradient is:
\begin{align}
    M^{\mathrm{CUC_{\infty}}}_{i}(x_e)
    =  - \frac{1}{n_k}
    \tup{
    \eta_{ik}\big(\vot_k\,t_e(x_e)+\pi_e\big)
    + 
    x_e^k \vot_k t'_e(x_e)
    } , \nonumber
\end{align}
where $k\in\arg\max_{h}\,\bar C_h$ with $\bar C_h=C_h/n_h$;
$\eta_{ik}$ is the direct impact (only for the worst group $k$), and the second term is the induced marginal congestion effect; both are scaled by its demand $n_k$.
As noted in Section~\ref{section: Comparability classes by invariance}, the limit $\rho \to \infty$ corresponds to the min–max formulation. In practice, this can be approximated by employing a sufficiently large value of $\rho$ and a smooth approximation to approach the minima.

\begin{algorithm}[tb]
\caption{Welfarism traffic control based on Frank-Wolfe algorithm}
\label{alg:fw}
\small
\begin{algorithmic}[1]
\Require Network $(\mathcal{V},\mathcal{E})$, demands $\requests$, tolls $s$, values of time $\{\vot_i\}$, objective selector $W^{\Phi}$, tolerance $\varepsilon$.
\State Initialize flow $x^{0}, t \leftarrow 0$
\While{$\big|W^{\Phi}(x^{t+1}) - W^{\Phi}(x^{t})\big| \ge \varepsilon$}
    \State $w' \leftarrow M^{\Phi}(x^t)$.
    \Comment{welfare margins} 
    \State $y^{t} \in \arg\max_{y} \sum_{e,i} y^i_e w^t_{i,e}$
    \Comment{auxiliary flow}
    \State $d^{t} \gets y^{t} - x^{t}$
    \Comment{descent direction}
    \State $\gamma \in \arg\max_{\gamma} W^{\Phi}\left((1-\gamma)x^{(t)} + \gamma d^{(t)}\right)$
    \State $x^{t+1} \gets (1-\gamma) x^{t} + \gamma d^{t}$
    \Comment{update flow}
    \EndWhile
\State \Return $x^{t}$
\end{algorithmic}
\vspace{0.5ex}
\footnotesize
\end{algorithm}

\section{Numerical experiments} \label{sec:Numerical experiments}

\subsection{Illustrative example}
\begin{figure}[tb]
    \centering
    \includegraphics[width=1\linewidth]{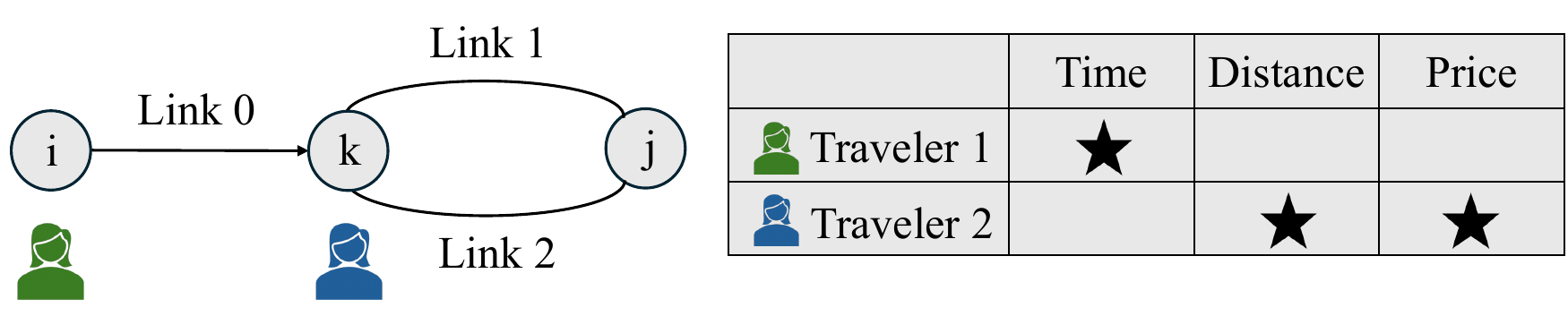}
    \caption{Two-traveler routing problem}
    \label{fig:Example}
\end{figure}

To illustrate the framework, consider a toy network in \ref{fig:Example}. 
The network has three links: two with fixed travel times and one (Link~2) whose travel time increases with flow due to congestion. 
There are two travelers with distinct preferences: Traveler~1 places greater weight on travel time, while Traveler~2 is more sensitive to distance and monetary cost. 
Each traveler’s total cost is a weighted sum of these components.
The parameters and link characteristics are provided in our GitHub repository\footnote{\url{https://github.com/mingjia-he/socialchoice_poa}}{}. 

We examine a toll on Link~2 as a simple policy lever to mitigate externalities, i.e., to reduce the inefficiency of selfish routing relative to the system optimum as captured by PoA. 
We compare two scenarios: \emph{Toll} \(=0\) and \emph{Toll} \(=4\). The resulting reference-adjusted PoA values under different comparability assumptions are summarized in Table~\ref{tab:example_result}.
\begin{table}[!ht]
    \centering
    \caption{PoA under comparability assumptions}
    \label{tab:example_result}
    \renewcommand{\arraystretch}{1.15}
    \begin{tabular}{cccc}
        \toprule
        \textbf{Toll rate} & $\mathbf{PoA^{CUC_0}}$ & $\mathbf{PoA^{CNC}}$ & $\mathbf{PoA^{CUC_{\infty}}}$ \\
        \midrule
        0   & 1.055 & 1.134 & 1.143 \\
        4   & 1.086 & 1.057 & 1.143 \\
        \bottomrule
    \end{tabular}
\end{table}

The evaluation of the tolling strategies depends on the assumed comparability of traveler utilities. 
When utilities are directly comparable—so that total system performance is the sum of Traveler~1’s and Traveler~2’s costs—\emph{Toll} \(=0\) is preferable to \emph{Toll} \(=4\). 
Under the CNC assumption, by contrast, \emph{Toll} \(=4\) yields the better outcome. 
If utilities are not comparable at all (max–min, \(\mathrm{CUC}_\infty\)), the two tolling schemes are indistinguishable.

In short, the ``appropriate'' toll depends critically on the chosen framework of utility comparability. 
Selecting an explicit and defensible comparability assumption is therefore essential for effective toll design. 
To demonstrate the practical implications at scale, we next examine a numerical experiment on the Zurich network.

\subsection{Real-world case study: Zurich network}
\begin{figure}[htb]
    \centering
    \includegraphics[width=0.8\linewidth]{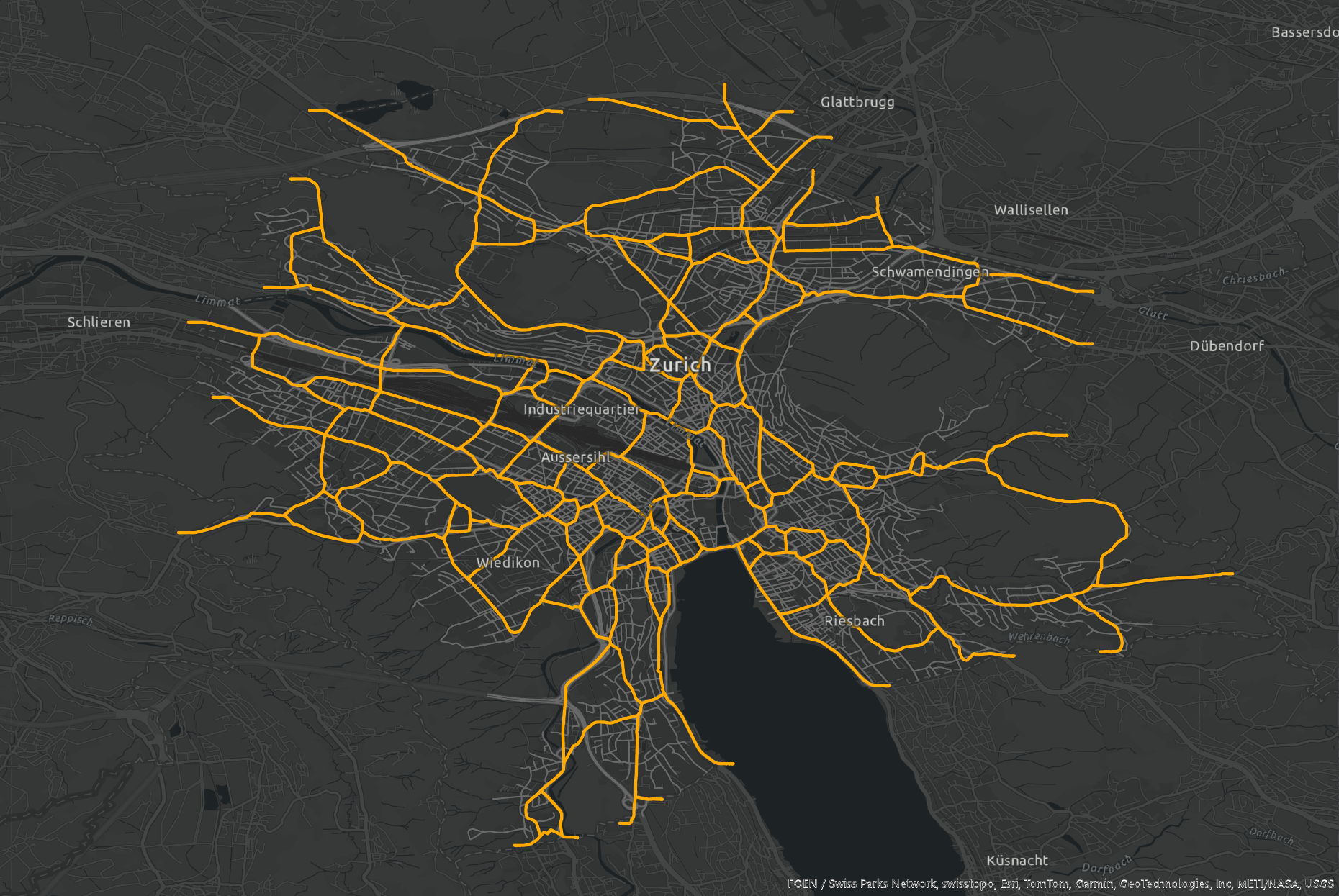}
    \caption{Study area: Zurich city.}
    \label{fig:enter-label}
\end{figure}

For this analysis, we obtained the road network topology of Zurich, Switzerland, from ~\cite{openstreetmap}. 
From this data, we extracted a simplified network representation, shown in Figure~\ref{fig:enter-label}, consisting of 188 nodes and 525 edges. 
Travel demand was derived from a one-day transportation simulation calibrated with population data from ~\cite{sbfsPortal}, resulting in 14,178 trips.

\begin{figure}[tb]
    \centering
    \includegraphics[width=0.95\linewidth]{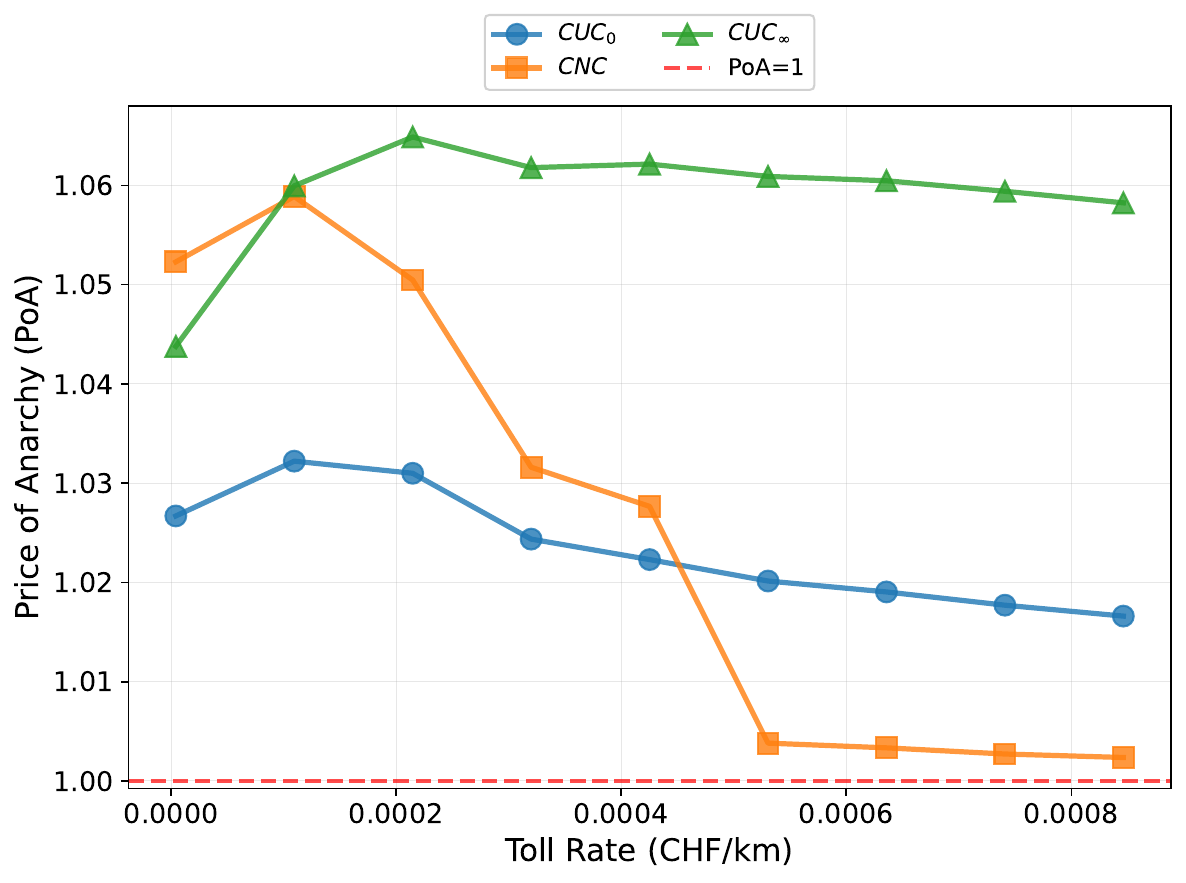}
    \caption{Price of Anarchy with various tolling rates.}
    \label{fig:toll_poa}
\end{figure}

The goal is to assess the influence of tolling rates on PoA and to examine how this evaluation result varies when adopting different assumptions about the comparability of user groups.
Figure~\ref{fig:toll_poa} presents the experimental results of the price of anarchy under various tolling rates.
%
A larger PoA indicates greater inefficiency in routing games under the given tolling rate.
The results indicate that, without tolling, the absolute PoA values differ depending on the assumed comparability condition.
As the tolling rate increases, the PoA tends to rise at first and then decline. Importantly, the rate at which system inefficiency reaches its maximum differs across the comparability assumptions.
When CUC$_0$ assumption is applied, the least efficient tolling rate is approximately $1.6\times 10^{-4}$ CHF/km, whereas under the CUC$_{\infty}$ assumption, it is around $2.2\times 10^{-4}$ CHF/km. 
Under the CUC$_0$ assumption, an increase in the tolling rate from $2\times 10^{-4}$ to $6\times 10^{-4}$ CHF/km leads to a substantial improvement in system efficiency, reflected by a reduction in PoA.
This trend is not observed under the other comparability assumptions.
Under the CUC$_{\infty}$ assumption, within the tolling range considered, increasing the toll rate can not lead to a more efficient outcome than having no tolling at all.
These findings highlight the critical importance of identifying the appropriate comparability assumption of user preferences and integrating it into the control mechanism. The assumed comparability structure directly influences the optimal policy decisions in traffic management and the resulting system efficiency.

\section{Conclusion}\label{sec: conclusion}
In this work, we introduce an axiomatic framework for the efficiency evaluation of socio-technical systems with selfish agents, considering the comparability among heterogeneous agents’ utilities. By enforcing invariance under specific utility transformations, we established a mapping between degrees of comparability and the appropriate social aggregators. We introduce an invariant PoA, which represents a robust efficiency measure that respects the informational structure of the game. 

The practical implication of this theoretical framework is demonstrated through our numerical analysis of the Zurich traffic network. The results reveal that the measured efficiency of tolling policies varies based on the underlying level of comparability, which underscores the importance of making comparability assumptions explicit to avoid misguided policy conclusions.
This finding highlights that robust decision-making requires robust metrics. With the invariant PoA, decision-makers can design mechanisms that are efficient and robust to the inherent subjectivity of agents’ preferences. 
Future work will focus on unifying different degrees of comparability within a single, continuous framework, and exploring its application in other decentralized domains, such as energy markets. Furthermore, we shall explore application of our framework to alternatives to the PoA, extending its to a wider set of equilibria notions (e.g. coarse-correlated equilibria), where the behavioral outcome can achieved naturally by no-regret learning.


\bibliography{references} 


\end{document}